\documentclass{article}
\usepackage{arxiv}
\usepackage{hyperref}
\usepackage{latexsym}
\usepackage{url}
\usepackage[utf8]{inputenc}
\usepackage{booktabs}
\usepackage{amsmath}
\usepackage{amsfonts}
\usepackage{amssymb}
\usepackage[inline]{enumitem}
\usepackage{enumitem}
\usepackage{mathbbol}
\usepackage[T1]{fontenc}
\usepackage{mathtools}
\usepackage{multirow}
\usepackage{natbib}
\usepackage{rotating}
\usepackage{subcaption}
\usepackage{graphicx}
\usepackage{etaremune}
\usepackage{authblk}
\usepackage{placeins}
\usepackage{tabularx}
\usepackage{tikz}

\newcommand{\vs}{\emph{vs. }}

\title{Overview of the TREC 2023 Product Product Search Track}
\author[1]{Daniel Campos}
\author[2]{Surya Kallumadi}
\author[3]{Corby Rosset}
\author[4]{ChengXiang Zhai}
\author[5]{Alessandro Magnani}
\affil[1]{Snowflake \\\texttt{\small daniel.campos@snowflake.com}}
\affil[2]{Lowes}
\affil[3]{Microsoft}
\affil[4]{University of Illinois Urbana-Champaign}
\affil[5]{Walmart}
\begin{document}
\maketitle

\begin{abstract}
This is the first year of the TREC Product search track. The focus this year was the creation of a reusable collection and evaluation of the impact of the use of metadata and multi-modal data on retrieval accuracy. This year we leverage the new product search corpus, which includes contextual metadata. Our analysis shows that in the product search domain, traditional retrieval systems are highly effective and commonly outperform general-purpose pretrained embedding models. Our analysis also evaluates the impact of using simplified and metadata-enhanced collections, finding no clear trend in the impact of the expanded collection. We also see some surprising outcomes; despite their widespread adoption and competitive performance on other tasks, we find single-stage dense retrieval runs can commonly be noncompetitive or generate low-quality results both in the zero-shot and fine-tuned domain.
\end{abstract}
\section{Introduction}
\label{sec:intro}
At TREC 2023, we hosted the first TREC Product Search Track, looking to create a reusable general benchmark for evaluating the performance of retrieval methods in the product search domain. We focus on providing a benchmark similar in scale and format to NQ \cite{Kwiatkowski2019NaturalQA}, or the Deep Learning Track \cite{craswell2021trec} but focused on product search. In providing a simple-to-use dataset, we believe broad experimentation using popular retrieval libraries \cite{Lin2021PyseriniAP} \cite{Gao2022TevatronAE} can lead to broad improvements in retrieval performance.\\
In this first year of the track, we created a novel collection based on the ESCI Product Re-ranking dataset \cite{Reddy2022ShoppingQD}, sampled novel queries, created enriched metadata in the form of additional text and images along with seeded evaluation results with a broad range of baseline runs to aid in collection reusability and to allow iteration and experimentation on the use of additional context. \\
Unlike previous product search corpora, the Product Search Track is multi-modal and has a large enough scale to explore the usage of neural retrieval methods. We observe somewhat surprising results using this scaled dataset and a wide variety of baseline runs. Single-stage retrieval models that leverage vector representations do not consistently outperform traditional retrieval methods such as BM25. Moreover, in the zero-shot setting, we find that larger vector-based models do not always beat their more minor variants, which is at odds with other evaluation corpora such as MTEB \cite{muennighoff2023mteb}. Finally, while additional metadata can improve retrieval performance at a macro level, extra information cannot guarantee performance. In evaluating per-query performance, we find that vector-based systems lose performance with the other metadata. Please see the participant papers for more insights about what we learned this year.
\section{Task description}
\label{sec:task}
The product search track has one task: product ranking. Within this task, various enriched datasets are opened to participants to allow them to enrich the collection as they see fit. Participants were allowed to submit up to three official runs. When submitting each run, participants indicated what external data, pretrained models, and other resources were used, as well as information on what style of the model was used.\\
In the ranking task, given a query, the participants were expected to retrieve a ranked list of products from the full collection based on the estimated likelihood that the product would meet the user's need. Participants could submit up to $100$ products per query for this end-to-end ranking task. \\
We first selected a subset of $200$ queries for judging in the pooling and judging process. Then NIST started judging these queries, throwing out queries without high disagreement or deemed un-judgable. If at least 50\%  of the judged products are relevant or there is no relevant product, the query is deemed un-judgable. This led to a judged test set of $186$ queries, which we compare the quality of runs. The track received 62 submissions to the passage ranking task, 39 of which were baseline runs that the track coordinators submitted. \\
Judgments were collected for each query product pair on a four-point scale:
\begin{etaremune}[start=3]
    \item \textbf{Perfectly relevant:} The product is exactly what the user wants.
    \item \textbf{Highly relevant:} The product could match the user query, but it may be a substitute for the original query intent. It may have a slightly different style, brand, or type, but a user would be satisfied if they received this product. 
    \item \textbf{Related:} The product seems related to the query but not the item the user seeks. Products in this category could complement the user's intended product. 
    \item \textbf{Irrelevant:} The product has nothing to do with the query.
\end{etaremune}
For binary metrics, we map judgment levels 3,2 to relevant and 1,0 to irrelevant.\\

\begin{figure}
    \centering
    \scalebox{0.5}{
    \includegraphics{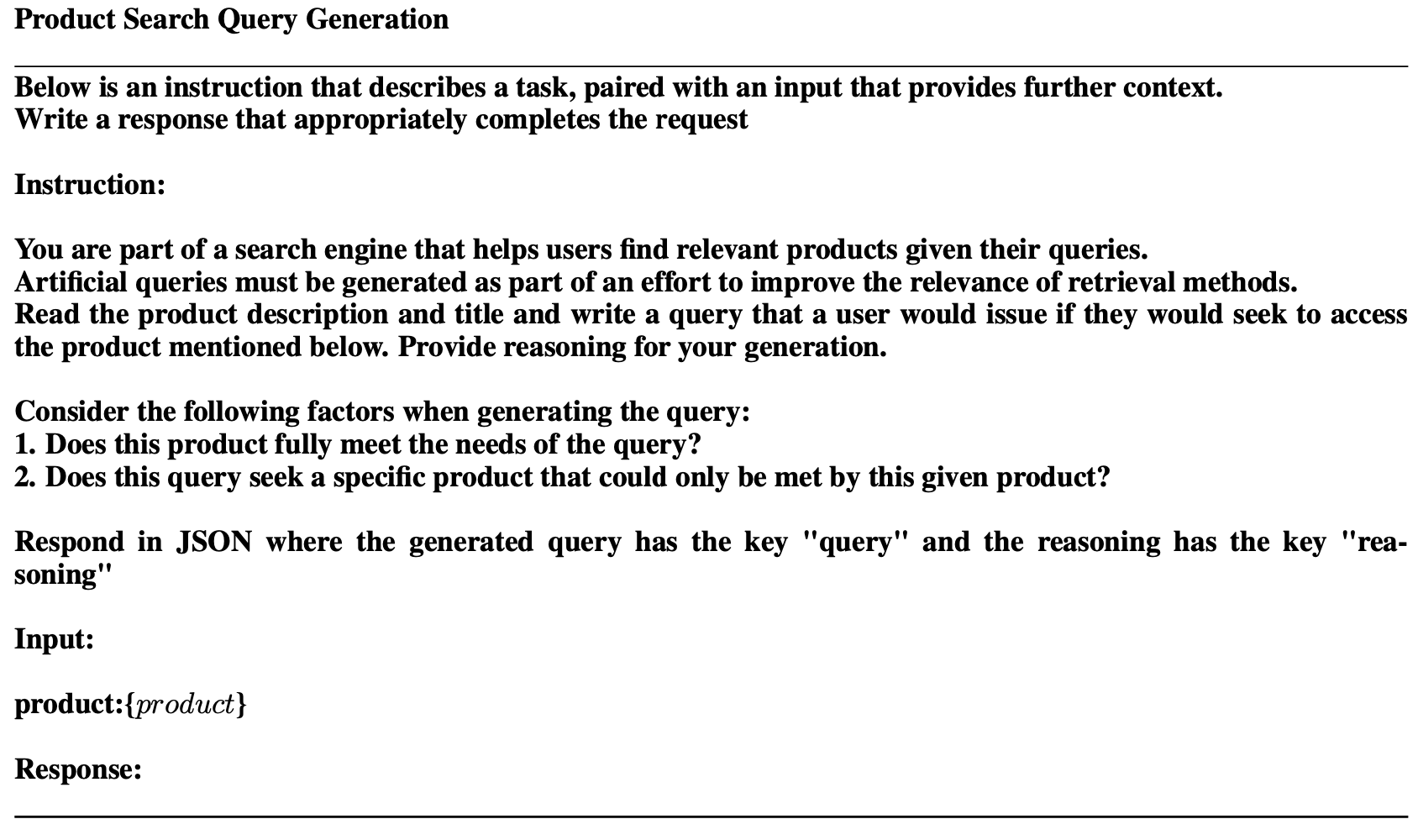}}
    \caption{Prompt used in our synthetic query generation on randomly selected product. The sampled product is included in the placeholders \{$product$\}.}
   \label{tab:query-prompt}
\end{figure}
\begin{figure}[!htb]
    \centering
    \includegraphics[width=1.0\linewidth]{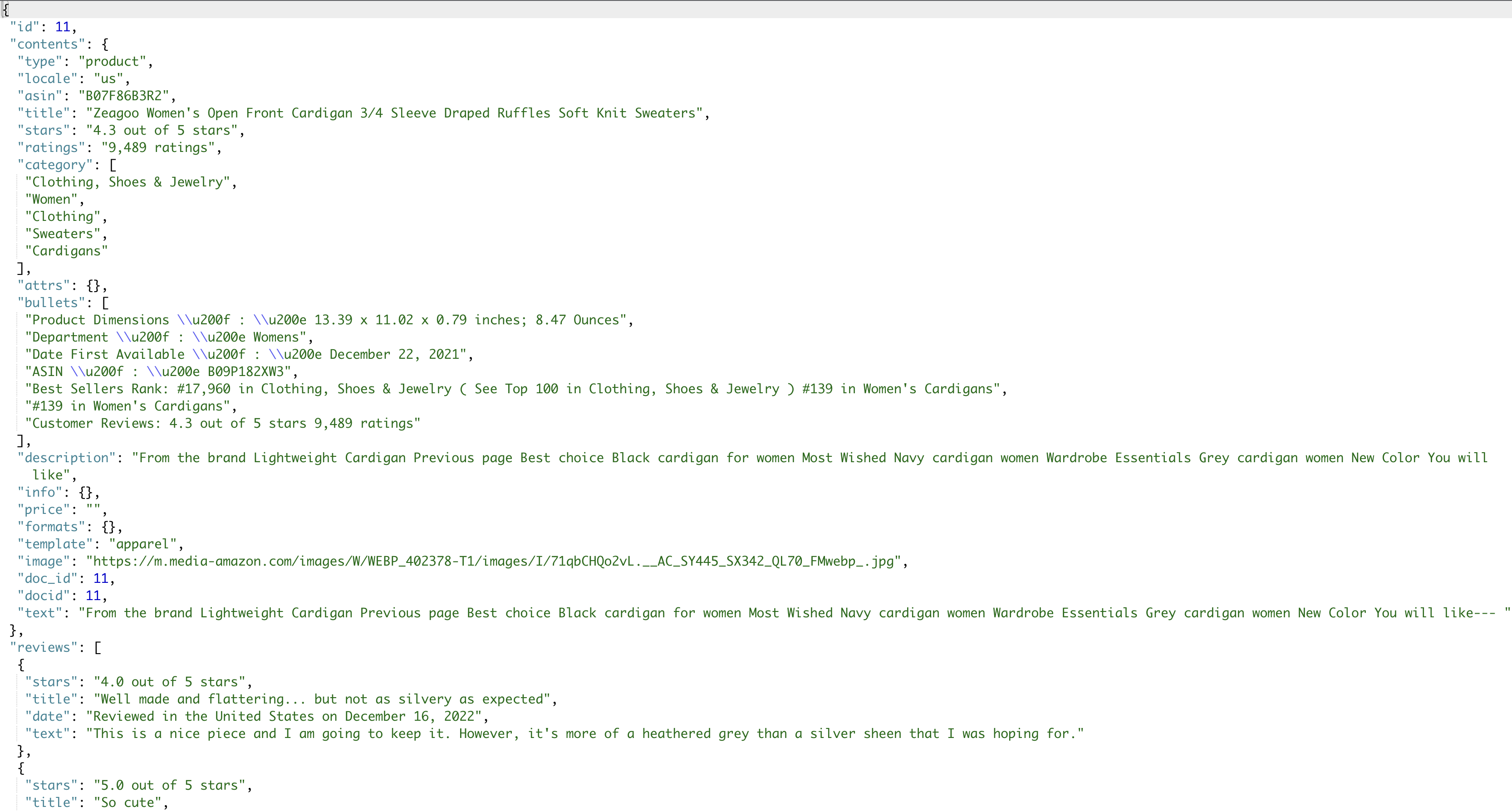}
    \caption{Each product in the collection contains basic information such as a title and product description along with contextual metadata, which includes attributes such as reviews, dimensions, etc.}
    \label{fig:example_product}
\end{figure}
\begin{figure}[!htb]
    \centering
    \includegraphics[]{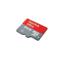}
    \includegraphics[]{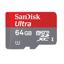}
    \includegraphics[]{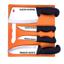}
    \includegraphics[]{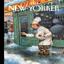}
    \includegraphics[]{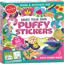}
    \includegraphics[]{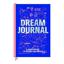}
    \caption{Some examples of product images. Some items have multiple images, while others have none.}
    \label{fig:example_img}
\end{figure}
The collection is based on the ESCI Shopping queries dataset \cite{Reddy2022ShoppingQD}. While this dataset is focused on improving product search, it lacks a clear end-to-end retrieval benchmark. Instead, the dataset includes a re-ranking task in which the top 40 results retrieved from the Amazon product corpus must be re-ranked for improved relevance. While this re-ranking task is quite important to the end-to-end performance of a product search engine, it does not allow for ample understanding of what impacts the performance of end-to-end retrieval in the product domain.  \\
Since there is no source of publicly available shopping queries, nor does the ESCI dataset have a publicly accessible test dataset, we created a new set of 998 evaluation queries leveraging GPT-4 and some heuristic-based sampling. For query generation, we leveraged GPT-4 along with Prompt, which built on the work of InPars \cite{Bonifacio2022InParsDA} \cite{Jeronymo2023InParsv2LL}, and we created 500 queries using the prompt shown in table \ref{tab:query-prompt}.\\
To avoid cases where this approach fails and to study how models perform with more typical product search queries with high keyword overlap, we generate queries by selecting sub-spans of product titles or descriptions. In generating queries with GPT-4, we aimed to create a reliable way of generating new and interesting queries for the collection, as we do not have a method to sample novel queries reliably. \\

\section{Datasets}
\label{sec:data}
This year, we leverage an enriched and filtered product search dataset based on the ESCI dataset \cite{Reddy2022ShoppingQD}. We will first describe the dataset and its generation before we describe how we adapted it to best suit the track. \\
\paragraph{Shopping Queries Dataset: A Large-Scale ESCI Benchmark for Improving Product Search}
The benchmark for improving product search, or ESCI, is a large-scale benchmarking dataset that focuses on subsets of product search use cases and provides frameworks from which improvement can be studied. Unlike other product retrieval datasets, the ESCI corpus contains English, Japanese, and Spanish queries.  \\
The dataset centers around three tasks important to the world of product search. It can be used to improve customer experience: Query-Product Ranking, Multi-class Product Classification, and Product Substitute Identification. For Query-Product Ranking: given a user-specified query and the top 40 products retrieved by a commercial product search engine, this task aims to rank the products to have more relevant products ranked above non-relevant products. For Multi-class Product Classification: given a query and a result of products, classify products into the following matches: Exact, substitute, complement, and irrelevant. For Product Substitute Identification: given a product and a list of potential substitutes, identify which could be substituted. \\
Within the three tasks, there are two variants of product collections, with the product ranking task using the smaller collection and the other tasks using a larger task. Given our focus on retrieval, we leverage the former. Within each task, there is a large training data set that contains query product pairs that have been annotated as exact match (E), substitute (S), complement (C), and irrelevant (I). The data contains the following fields: example id, query, query id, product id, product locale, ESCI label, small version, large version, split, product title, product description, product bullet point, product brand, product color, and source. \\
The smaller ranking dataset has 48,300 unique queries and 1,118,011 relevance judgments. The data sets are stratified into train, dev, and test, of which only the labels for the train and dev have been released publicly. On average, each query has 20 judgments for English and 28 for other languages. \\
\begin{table}[!ht]
    \centering
    \begin{tabular}{|l|l|l|}
    \hline
        Item & Instances & Notes \\ \hline
        Collection & 1,661,907 & 90+\% of products have at least 1 image \\ \hline
        Train Queries & 30,734 & Train + Dev \\ \hline
        Train QREL & 392,119.00 & N/A \\ \hline
        2023 Test Queries & 926 & N/A \\ \hline
        2023 Test Queries (Judged) & 182 & N/A \\ \hline
    \end{tabular}
    \caption{High-level statistics on the size of the collection and queries of the TREC Product Search 2023 Collection}
    \label{tab:dataset-stat}
\end{table}
\paragraph{Product Search Track Corpus}
While the full ESCI dataset is multilingual and features over 3 million items, we narrowed our focus to English only. We attempt to enrich the dataset with additional Metadata and images for these English products as we believe this can be very important for product search. The ESCI dataset is focused on text information that ignores any behavioral, categorical, visual, or numerical features that can be used for ranking. Product metadata enrichment improves product representations by including additional helpful information such as reviews, attributes such as size and color, and categorical ordering from extracted Metadata from Amazon's online catalog \footnote{Metadata was extracted from https://github.com/shuttie/esci-s/}.\\
Figure \ref{fig:example_product} shows an example product with its additional Metadata. We crawled and removed images for each product using the ASIN from this enrichment. Product images contain one to ten thumbnail-size images for a given product, which were shuffled from Amazon and joined with the ESCI dataset. Since these images are extracted from product thumbnails, each image is only 64x64, which allows the entire collection to be relatively small. Some product image examples can be found in figure \ref{fig:example_img}. Numerical details on the collection can be found in table \ref{tab:dataset-stat}
\section{Results and analysis}
\label{sec:result}
\begin{table}
    \centering
    \caption{TREC 2023 Product Search Track run submission statistics.}
    \begin{tabular}{lrr}
    \hline
    \hline
        & \textbf{All Groups} & \textbf{Coordinator Baselines} \\
        \hline
        Number of groups & 4 & 1 \\
        Number of total runs & 62 & 39 \\
        \hline
        \hline
    \end{tabular}
    \label{tbl:runs-by-type}
\end{table}
\begin{figure}[!htb]
    \centering
    \includegraphics[width=1.0\linewidth]{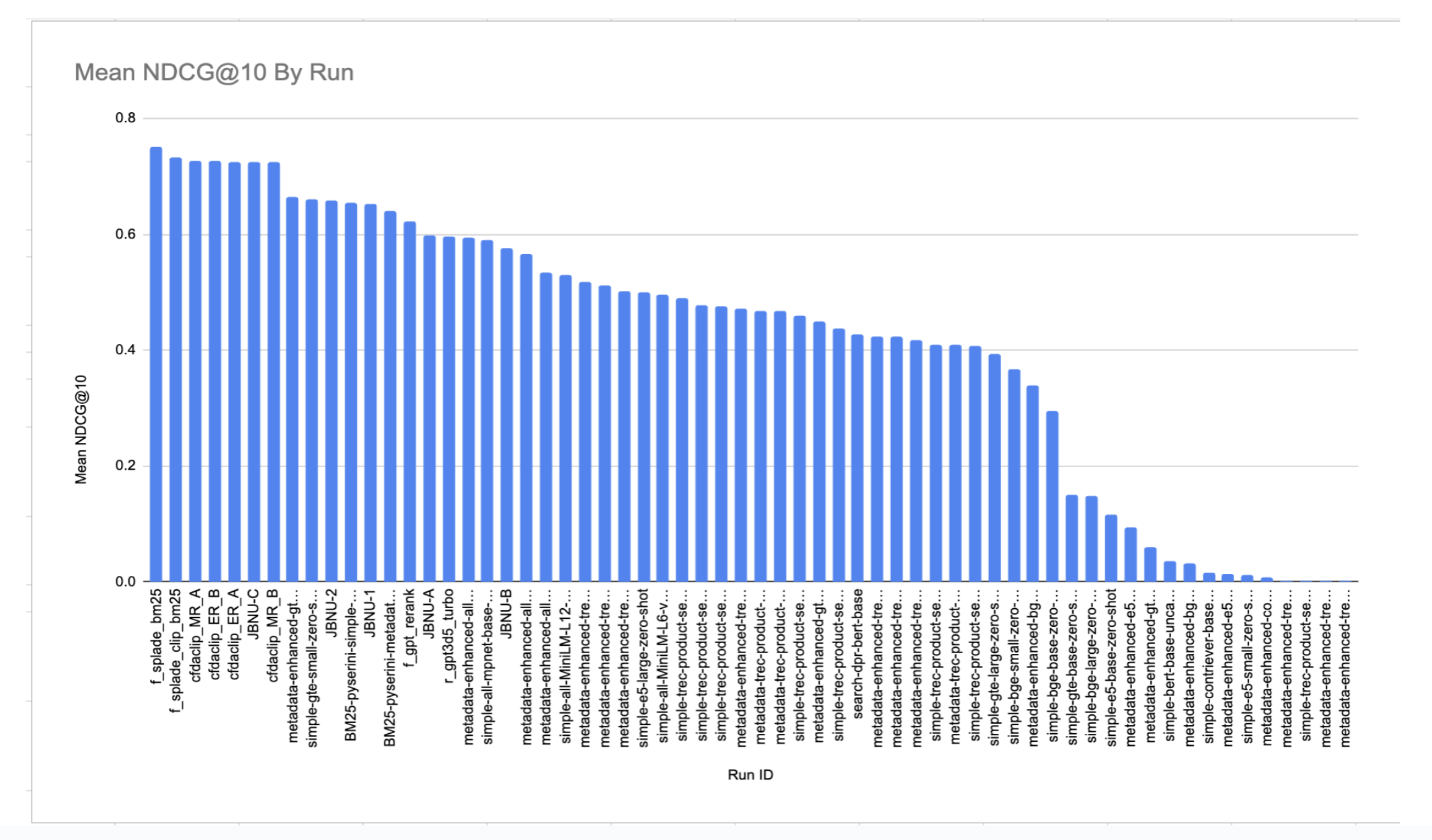}
    \caption{Relative system ordering based on mean NDC@10}
    \label{fig:ndcg@10summary}
\end{figure}
\paragraph{Submitted runs}
A total of $4$ groups participated in the TREC 2023 Product Search Track, including baseline runs submitted by the track coordinators. Across all groups, we received $62$ run submissions, including $49$ baseline runs.
Table~\ref{tbl:runs-by-type} and figure \ref{fig:ndcg@10summary} summarizes the submission statistics for this year's track. This set of runs led to $182$ evaluated queries, which we believe will likely make this a highly reusable collection apt for future experimentation.\\
\begin{table}[!htb]
\caption{Summary of results for all runs.}
\scriptsize
\centering
        \begin{tabular}{|l|l|l|l|l|l|l|}
    \hline
    \toprule
        Run & Group & InfAP & NDCG @10 & NDCG @100 & R@10 & R@100 \\
        \midrule
        f\_splade\_bm25 & F & 0.6068 & 0.7505 & 0.7244 & 0.4919 & 0.8015 \\ \hline
        f\_splade\_clip\_bm25 & F & 0.5731 & 0.7327 & 0.7143 & 0.4739 & 0.8001 \\ \hline
        cfdaclip\_MR\_A & JBNU & 0.5910 & 0.7257 & 0.7019 & 0.4766 & 0.7857 \\ \hline
        cfdaclip\_ER\_B & JBNU & 0.5905 & 0.7256 & 0.7010 & 0.4766 & 0.7862 \\ \hline
        cfdaclip\_ER\_A & JBNU & 0.5902 & 0.7252 & 0.7008 & 0.4765 & 0.7840 \\ \hline
        JBNU-C & JBNU & 0.5885 & 0.7251 & 0.7074 & 0.4700 & 0.7870 \\ \hline
        cfdaclip\_MR\_B & JBNU & 0.5903 & 0.7251 & 0.7010 & 0.4765 & 0.7859 \\ \hline
        metadata-enhanced-gte-small-zero-shot & Baselines & 0.4955 & 0.6647 & 0.6500 & 0.4363 & 0.7416 \\ \hline
        simple-gte-small-zero-shot & Baselines & 0.4818 & 0.6612 & 0.6492 & 0.4375 & 0.7372 \\ \hline
        JBNU-2 & JBNU & 0.4792 & 0.6583 & 0.6208 & 0.4359 & 0.7272 \\ \hline
        BM25-pyserini-simple-collection & Baselines & 0.4769 & 0.6540 & 0.6148 & 0.4287 & 0.7241 \\ \hline
        JBNU-1 & JBNU & 0.4828 & 0.6531 & 0.6185 & 0.4092 & 0.7272 \\ \hline
        BM25-pyserini-metadata-collection & Baselines & 0.4729 & 0.6408 & 0.6160 & 0.4254 & 0.7272 \\ \hline
        f\_gpt\_rerank & F & 0.4673 & 0.6225 & 0.6599 & 0.3765 & 0.8001 \\ \hline
        JBNU-A & JBNU & 0.4500 & 0.5989 & 0.5607 & 0.3772 & 0.6636 \\ \hline
        r\_gpt3d5\_turbo & r & 0.4174 & 0.5950 & 0.5889 & 0.3806 & 0.7272 \\ \hline
        metadata-enhanced-all-mpnet-base-v2-zero-shot & Baselines & 0.4144 & 0.5937 & 0.5541 & 0.3862 & 0.6512 \\ \hline
        simple-all-mpnet-base-v2-zero-shot & Baselines & 0.4000 & 0.5895 & 0.5508 & 0.3806 & 0.6348 \\ \hline
        JBNU-B & JBNU & 0.4349 & 0.5763 & 0.5380 & 0.3580 & 0.6339 \\ \hline
        metadata-enhanced-all-MiniLM-L12-v2-zero-shot & Baselines & 0.3844 & 0.5660 & 0.5309 & 0.3821 & 0.6558 \\ \hline
        metadata-enhanced-all-MiniLM-L6-v2-zero-shot & Baselines & 0.3688 & 0.5328 & 0.5164 & 0.3654 & 0.6415 \\ \hline
        simple-all-MiniLM-L12-v2-zero-shot & Baselines & 0.3483 & 0.5288 & 0.5161 & 0.3502 & 0.6365 \\ \hline
        metadata-enhanced-trec-product-search-gte-small & Baselines & 0.3520 & 0.5168 & 0.5101 & 0.3443 & 0.5859 \\ \hline
        metadata-enhanced-trec-product-search-e5-small-v2 & Baselines & 0.3488 & 0.5119 & 0.5082 & 0.3481 & 0.6096 \\ \hline
        metadata-enhanced-trec-product-search-gte-base & Baselines & 0.3423 & 0.5009 & 0.5004 & 0.3400 & 0.5895 \\ \hline
        simple-e5-large-zero-shot & Baselines & 0.3339 & 0.4998 & 0.4490 & 0.3428 & 0.5537 \\ \hline
        simple-all-MiniLM-L6-v2-zero-shot & Baselines & 0.3261 & 0.4952 & 0.4924 & 0.3334 & 0.6099 \\ \hline
        simple-trec-product-search-gte-small & Baselines & 0.3194 & 0.4901 & 0.4902 & 0.3080 & 0.5692 \\ \hline
        simple-trec-product-search-gte-base & Baselines & 0.3067 & 0.4777 & 0.4813 & 0.3123 & 0.5676 \\ \hline
        simple-trec-product-search-all-miniLM-L12-v2 & Baselines & 0.3060 & 0.4763 & 0.4589 & 0.3100 & 0.5351 \\ \hline
        metadata-enhanced-trec-product-search-bge-small-en & Baselines & 0.3193 & 0.4721 & 0.4708 & 0.3012 & 0.5565 \\ \hline
        metadata-trec-product-search-all-miniLM-L12-v2 & Baselines & 0.3129 & 0.4681 & 0.4603 & 0.3081 & 0.5581 \\ \hline
        metadata-trec-product-search-all-miniLM-L6-v2 & Baselines & 0.3144 & 0.4673 & 0.4675 & 0.3181 & 0.5528 \\ \hline
        simple-trec-product-search-all-miniLM-L6-v2 & Baselines & 0.3008 & 0.4591 & 0.4599 & 0.2931 & 0.5429 \\ \hline
        metadata-enhanced-gte-large-zero-shot & Baselines & 0.2503 & 0.4501 & 0.4103 & 0.2698 & 0.4978 \\ \hline
        simple-trec-product-search-bge-small-en & Baselines & 0.2726 & 0.4379 & 0.4328 & 0.2741 & 0.5080 \\ \hline
        search-dpr-bert-base & Baselines & 0.2648 & 0.4272 & 0.4333 & 0.2796 & 0.5068 \\ \hline
        metadata-enhanced-trec-product-search-e5-base-v2 & Baselines & 0.2703 & 0.4242 & 0.4118 & 0.2793 & 0.5148 \\ \hline
        metadata-enhanced-trec-product-search-bge-base-en & Baselines & 0.2709 & 0.4237 & 0.4165 & 0.2938 & 0.4955 \\ \hline
        metadata-enhanced-trec-product-search-dpr-bert & Baselines & 0.2636 & 0.4165 & 0.4276 & 0.2774 & 0.5208 \\ \hline
        simple-trec-product-search-all-mpnet-base-v2 & Baselines & 0.2377 & 0.4090 & 0.4013 & 0.2507 & 0.4747 \\ \hline
        metadata-trec-product-search-all-mpnet-base-v2 & Baselines & 0.2611 & 0.4089 & 0.4118 & 0.2643 & 0.5006 \\ \hline
        simple-trec-product-search-bge-base-en & Baselines & 0.2448 & 0.4064 & 0.4027 & 0.2628 & 0.4728 \\ \hline
        simple-gte-large-zero-shot & Baselines & 0.2146 & 0.3930 & 0.3654 & 0.2319 & 0.4294 \\ \hline
        simple-bge-small-zero-shot & Baselines & 0.1898 & 0.3680 & 0.3475 & 0.2059 & 0.4188 \\ \hline
        metadata-enhanced-bge-base-en-zero-shot & Baselines & 0.1919 & 0.3396 & 0.3290 & 0.2211 & 0.4301 \\ \hline
        simple-bge-base-zero-shot & Baselines & 0.1178 & 0.2948 & 0.2458 & 0.1479 & 0.2664 \\ \hline
        simple-gte-base-zero-shot & Baselines & 0.0522 & 0.1493 & 0.1131 & 0.0581 & 0.0965 \\ \hline
        simple-bge-large-zero-shot & Baselines & 0.0498 & 0.1486 & 0.1056 & 0.0537 & 0.0787 \\ \hline
        simple-e5-base-zero-shot & Baselines & 0.0439 & 0.1168 & 0.0938 & 0.0527 & 0.0894 \\ \hline
        metadata-enhanced-e5-base-v2-zero-shot & Baselines & 0.0276 & 0.0936 & 0.0861 & 0.0375 & 0.1021 \\ \hline
        metadata-enhanced-gte-base-zero-shot & Baselines & 0.0285 & 0.0604 & 0.0614 & 0.0333 & 0.0940 \\ \hline
        simple-bert-base-uncased-zero-shot & Baselines & 0.0074 & 0.0352 & 0.0294 & 0.0100 & 0.0374 \\ \hline
        metadata-enhanced-bge-large-en-zero-shot & Baselines & 0.0101 & 0.0323 & 0.0287 & 0.0108 & 0.0314 \\ \hline
        simple-contriever-base-zero-shot & Baselines & 0.0049 & 0.0159 & 0.0159 & 0.0062 & 0.0372 \\ \hline
        metadata-enhanced-e5-small-v2-zero-shot & Baselines & 0.0102 & 0.0142 & 0.0116 & 0.0090 & 0.0126 \\ \hline
        simple-e5-small-zero-shot & Baselines & 0.0071 & 0.0113 & 0.0098 & 0.0089 & 0.0130 \\ \hline
        metadata-enhanced-contriever-base-msmarco & Baselines & 0.0022 & 0.0081 & 0.0108 & 0.0026 & 0.0332 \\ \hline
        metadata-enhanced-trec-product-search-bge-large-en & Baselines & 0.0000 & 0.0021 & 0.0008 & 0.0000 & 0.0000 \\ \hline
        simple-trec-product-search-gte-large & Baselines & 0.0000 & 0.0015 & 0.0007 & 0.0001 & 0.0001 \\ \hline
        metadata-enhanced-trec-product-search-e5-large-v2 & Baselines & 0.0000 & 0.0015 & 0.0007 & 0.0000 & 0.0000 \\ \hline
        metadata-enhanced-trec-product-search-gte-large & Baselines & 0.0000 & 0.0011 & 0.0006 & 0.0000 & 0.0000 \\ \hline
\bottomrule
\end{tabular}
\label{tab:full-result-summary}
\end{table}
\begin{table}[!ht]
    \centering
    \tiny
    \begin{tabular}{|l|l|l|}
    \hline
        Query & Max-Mean Gap & Systems Score 0 \\ \hline
        Elite (Elite) Volano/drivo/Kura For Body 329770001 & 0.9781 & 96.77\% \\ \hline
        small measuring rice bin Asvel & 0.8037 & 70.97\% \\ \hline
        Switch protective film Japanese glass blue light reduction water repellent anti-fingerprint & 0.7968 & 77.42\% \\ \hline
        Elegant satin floral lace ribbon lingerie set & 0.7264 & 59.68\% \\ \hline
        onlypuff Pocket Shirts for Women Casual & 0.6750 & 27.42\% \\ \hline
        Ekouaer Long Nightgown,Women's Loungewear Short Sleeve & 0.6590 & 33.87\% \\ \hline
        Lugz Women's Empire Hi Wvt Fashion Boot & 0.6559 & 24.19\% \\ \hline
        Dinosaur Pee Pee Teepee Wee & 0.6553 & 41.94\% \\ \hline
        10th birthday decorations for girl & 0.6548 & 19.35\% \\ \hline
        juDanzy kids knee high tube socks with grips & 0.6461 & 58.06\% \\ \hline
        Matching Delivery Robe and Swaddle Blanket & 0.6429 & 41.94\% \\ \hline
        UCGOU Bubble Mailers 7.25x12 Inch Teal 25 Pack & 0.6277 & 24.19\% \\ \hline
        Cicy Bell Women's Sunflower & 0.6211 & 62.90\% \\ \hline
        Canomo Lamp Light Kit Make a & 0.6209 & 19.35\% \\ \hline
        5L matte black rectangular trash can with soft close lid and anti-bag slip liner for bathroom or kitchen & 0.6206 & 50.00\% \\ \hline
        Women's UPF 50+ cotton linen bucket sun hat beige small & 0.6163 & 24.19\% \\ \hline
        Small breed wet dog food Hill's Science Diet Chicken \& Barley Recipe & 0.6038 & 40.32\% \\ \hline
        DKB Evian Jetted Whirlpool & 0.6031 & 27.42\% \\ \hline
        OtterBox Symmetry Disney Princess Mulan iPhone Xs iPhone X case & 0.6002 & 45.16\% \\ \hline
        Stars in the Desert book & 0.5958 & 58.06\% \\ \hline
        DC Collectibles Batman Arkham Origins & 0.5930 & 35.48\% \\ \hline
        fall sunflower pumpkin placemats set of 6 cotton linen washable table mats & 0.5853 & 27.42\% \\ \hline
        300 piece jigsaw puzzle Kitchen Memories by Steve Crisp & 0.5821 & 30.65\% \\ \hline
        Marvel Avengers Endgame Gauntlet T-Shirt & 0.5816 & 20.97\% \\ \hline
        Pahajim Women Fashion Purses Handbags Shoulder Tote Bags Top Handle Satchel & 0.5811 & 46.77\% \\ \hline
        Mai Puru Endo Mai's First Photo Collection & 0.5806 & 66.13\% \\ \hline
        girls princess dress up costume headband accessories & 0.5764 & 19.35\% \\ \hline
        Nanatang Badflower Logo Men's Long Sleeve Sweatshirt's & 0.5756 & 14.52\% \\ \hline
        HAPY SHOP 80 Pcs Silver Alligator Hair & 0.5705 & 45.16\% \\ \hline
        VERYKE L-Shaped sectional sofa chenille fabric golden legs living room & 0.5689 & 27.42\% \\ \hline
        ZINUS Owen Wood Platform & 0.5667 & 29.03\% \\ \hline
        Xperia 10 II Blue Light Cut Glass Film Asahi Japanese Ultra Thin Anti-Bubble Anti-Fingerprint & 0.5630 & 67.74\% \\ \hline
        Acer V6 V196LB 19" HD & 0.5619 & 50.00\% \\ \hline
    \end{tabular}
    \caption{Per query gap between Mean NDCG@10 and Max NDCG and the \% of retrieval systems with a NDCG@10 of zero.  }
    \label{tab:per-query}
\end{table}
This year, we had fewer participating groups than we hoped for compared to similar tracks (Deep learning had $15$ groups in 2019, $25$ in 2020, and $19$ in 2021). We believe this might indicate the broader saturation of the IR community by large-scale datasets focused on single-stage retrieval via neural language models. \\
\paragraph{Overall results}
Table~\ref{tab:full-result-summary} presents a standard set of relevance quality metrics for product search ranking runs. Reported metrics include Normalized Discounted Cumulative Gain (NDCG)~\citep{JK2002} at depth 10 and 100, Recall at depth 10 and 100, and the Infrared Average Precision (InfAP) \citep{Yilmaz2006EstimatingAP}. Results represent the mean score across the $182$ queries that NIST assessed, and scores are computed using TRECEVAL. None of these results leverage the existing development portion of ESCI or the unreleased eval set. In subsequent discussions, we employ NDCG@10 as our primary evaluation metric to analyze the ranking quality produced by different methods.\\
To analyze how different approaches perform in the high recall domain, we employ recall at 100 (R@100), which compares how often the positive set is present in the top 100 candidates even if they are not often ranked highly. In the product search, users often add sorting and filtering forms via price, size, color, etc. When a user removes portions of the ranked candidate set, the recall of a larger filter set becomes highly important. \\
Looking at the results in table \ref{tab:full-result-summary}, we see clear gains from hybrid retrieval systems that leverage multiple retrievers to improve performance (f\_splade\_bm25,cfdaclip\_MR\_A). We further see that in this domain, there is a consistent and effective performance for traditional retrieval methods such as BM25, which is one of the top-performing systems despite a lack of collection optimization. \\
When we evaluate specific queries as shown in table \ref{tab:per-query}, we find that there are some queries, such as \textit{Elite (Elite) Volano/drivo/Kura For Body 329770001} or \textit{small measuring rice bin Asvel} or \textit{Dinosaur Pee Pee Teepee Wee}, where one or a few retrieval systems have high NDCG@10 scores. In contrast, most systems have scores of 0. Each of these queries is looking for specific items, and surprisingly, the systems that excel at spear-fishing each product are inconsistent across questions. 
\paragraph{Performance on long \vs short queries}
To better understand the variance between variations in query length, we stratified the queries based on query length to analyze whether system ordering depends on query length. We study this because shorter queries, such as \textit{Google Wi-Fi System Mesh}, tend to be more broad in the world of product search. In contrast, longer queries, such as \textbf{21x21 beige sun shade sail patio UV protection outdoor backyard}, focus on finding specific products. We stratify queries by setting queries with 7 or more words that are too long and everything else assorted. This leaves 182 total queries, 81 short queries, and 101 long queries. When we use these stratified sets of queries, we find high Kendall tau with 0.93442 (7.13e-27) and  0.9640 (1.70e-28) for short and long, respectively. This agreement is surprisingly high as we see a large variation in NDCG as shown in table \ref{tab:query-length} where, for example, with BM25, the NDCG@10 from long to short queries is > 10\% relative move. \\
\paragraph{Metadata \vs Simple Collection}
\begin{table}[!htb]
    \centering
    \begin{tabular}{|l|l|l|l|l|}
    \hline
        Run & Zero Shot & NDCG @10 (metadata) & NDCG @10 (simple) & Impact \\ \hline
        BM25 & Y & 0.6408 & 0.6540 & -0.0133 \\ \hline
        all-MiniLM-L12-v2 & Y & 0.5660 & 0.5288 & 0.0371 \\ \hline
        all-MiniLM-L6-v2 & Y & 0.5328 & 0.4952 & 0.0376 \\ \hline
        all-mpnet-base-v2 & Y & 0.5937 & 0.5895 & 0.0043 \\ \hline
        bge-base-en & Y & 0.3396 & 0.2948 & 0.0448 \\ \hline
        bge-large-en & Y & 0.0323 & 0.1486 & -0.1163 \\ \hline
        contriever-base & Y & 0.0081 & 0.0159 & -0.0079 \\ \hline
        e5-base-v2 & Y & 0.0936 & 0.1168 & -0.0232 \\ \hline
        e5-small-v2 & Y & 0.0142 & 0.0113 & 0.0029 \\ \hline
        gte-base & Y & 0.0604 & 0.1493 & -0.0889 \\ \hline
        gte-large & Y & 0.4501 & 0.3930 & 0.0571 \\ \hline
        gte-small & Y & 0.6647 & 0.6612 & 0.0035 \\ \hline
        bge-base-en & N & 0.4237 & 0.4064 & 0.0174 \\ \hline
        bge-small-en & N & 0.4721 & 0.4379 & 0.0342 \\ \hline
        Bert-base & N & 0.4165 & 0.4272 & -0.0107 \\ \hline
        gte-base & N & 0.5009 & 0.4777 & 0.0233 \\ \hline
        gte-large & N & 0.0011 & 0.0015 & -0.0005 \\ \hline
        gte-small & N & 0.5168 & 0.4901 & 0.0267 \\ \hline
        all-miniLM-L12-v2 & N & 0.4681 & 0.4763 & -0.0082 \\ \hline
        all-miniLM-L6-v2 & N & 0.4673 & 0.4591 & 0.0082 \\ \hline
        all-mpnet-base-v2 & N & 0.4089 & 0.4090 & -0.0001 \\ \hline
    \end{tabular}
    \caption{NDCG@10 performance of retrieval methods using the simple collection and metadata enhanced collection.}
    \label{tab:meta-data-ndg}
\end{table}
\begin{table}[!htb]
    \centering
    \begin{tabular}{|l|l|l|l|l|}
    \hline
        Run & Zero Shot & R@100 (Metadata) & R@100 (Simple) & Impact \\ \hline
        BM25 & Y & 0.7272 & 0.7241 & 0.0032 \\ \hline
        all-MiniLM-L12-v2 & Y & 0.6558 & 0.6365 & 0.0193 \\ \hline
        all-MiniLM-L6-v2 & Y & 0.6415 & 0.6099 & 0.0316 \\ \hline
        all-mpnet-base-v2 & Y & 0.6512 & 0.6348 & 0.0163 \\ \hline
        bge-base-en & Y & 0.4301 & 0.2664 & 0.1637 \\ \hline
        bge-large-en & Y & 0.0314 & 0.0787 & -0.0473 \\ \hline
        contriever-base & Y & 0.0332 & 0.0372 & -0.0040 \\ \hline
        e5-base-v2 & Y & 0.1021 & 0.0894 & 0.0127 \\ \hline
        e5-small-v2 & Y & 0.0126 & 0.0130 & -0.0005 \\ \hline
        gte-base & Y & 0.0940 & 0.0965 & -0.0026 \\ \hline
        gte-large & Y & 0.4978 & 0.4294 & 0.0684 \\ \hline
        gte-small & Y & 0.7416 & 0.7372 & 0.0044 \\ \hline
        bge-base-en & N & 0.4955 & 0.4728 & 0.0227 \\ \hline
        bge-small-en & N & 0.5565 & 0.5080 & 0.0485 \\ \hline
        Bert-base & N & 0.5208 & 0.5068 & 0.0140 \\ \hline
        gte-base & N & 0.5895 & 0.5676 & 0.0219 \\ \hline
        gte-large & N & 0.0000 & 0.0001 & -0.0001 \\ \hline
        gte-small & N & 0.5859 & 0.5692 & 0.0167 \\ \hline
        all-miniLM-L12-v2 & N & 0.5581 & 0.5351 & 0.0230 \\ \hline
        all-miniLM-L6-v2 & N & 0.5528 & 0.5429 & 0.0099 \\ \hline
        all-mpnet-base-v2 & N & 0.5006 & 0.4747 & 0.0259 \\ \hline
    \end{tabular}
    \caption{Recall@100 performance of retrieval methods using the simple collection and metadata enhanced collection.}
    \label{tab:meta-data-r}
\end{table}
To understand the impact of enriching the collection with textual metadata in our baselines, we provided runs that use the simple collection and enhanced metadata. Tables \ref{tab:meta-data-ndg} and \ref{tab:meta-data-r} provide detailed results on the impact of using metadata vs simple data across some baseline runs. Based on this data, we note that when focused on NDCG@10, the introduction of metadata is somewhat of a mixed message. Some models see benefit from the additional information, and others see losses. \\
Despite this variability, our impact tends to be relatively small, as good retrievers models know an effect of less than 5\%. When focused on recall, the message changes, as most models see an improvement in recall by using metadata. However, like the impact on the top ten, recall improvements are minor, with a few exceptions. Across both metrics, we do not see any impact trend related to fine-tuned or zero-shot models.
\paragraph{Finetune \vs Zero-Shot}
As part of our baselines, we evaluated a set of naive baselines where we finetune in a single form and compare the impact of fine-tuning across runs. We leverage the Tevatron library and follow the training procedure from the NQ implementation \footnote{https://github.com/texttron/tevatron/blob/main/examples/}. We train each model for 40 epochs using 4 A100 with a batch size of 128, cross-device negatives, and learning rates of 1e-5,2e-5,3e-5,5e-5, and 1e-4, selecting the model that had the lowest validation loss at the end. These runs are not meant to be highly optimized fine-tuning runs but general explorations on the impact of fine-tuning. As shown in tables \ref{tab:finetune-zero-n} and \ref{tab:finetune-zero-r}, we see a consistent trend that the larger models suffer from fine-tuning (indicating the fine-tuning recipe was incorrect), but smaller models see large gains, in some cases going from completely unusable to highly competitive. \\
\FloatBarrier
\section{Conclusion}
\label{sec:conclusion}
This is the first year of the TREC Product Search Track. In our first year, we aimed to create a complete collection that could reliably evaluate the performance of different retrieval methods in the product search domain. In creating this initial collection, we introduce forms of collection enrichment via product images and metadata. Using this collection, we create a large set of artificial queries using a large language model and evaluate system performance using pooled judgment via query-product relevant judgments. \\
While this year's participation was light, we believe that the strong fundamentals of this collection pave the way to broader experimentation. This report summarizes the product search track's creation, the systems' high-level performance, and the perceived impact of additional metadata. 
\section*{Acknowledgement}
We thank Noveen Sachdeva, who helped find, extract, and process the images for this track. 

\bibliographystyle{plainnat}
\bibliography{bibtex}
\newpage
\appendix
\section{Per Query Performance}
\label{sec:appendix}
\begin{table}[]
    \centering
    \tiny
    \scalebox{0.8}{
    \begin{tabular}{|l|l|l|l|l|}
    \hline
        Query & Max & Mean & Gap Max vs Mean & Systems with NDCG@10 = 0 \\ \hline
        Elite (Elite) Volano/drivo/Kura For Body 329770001 & 1.0000 & 0.0219 & 0.9781 & 96.77\% \\ \hline
        small measuring rice bin Asvel & 1.0000 & 0.1963 & 0.8037 & 70.97\% \\ \hline
        Switch protective film Japanese glass blue light reduction water repellent anti-fingerprint & 0.9056 & 0.1088 & 0.7968 & 77.42\% \\ \hline
        Elegant satin floral lace ribbon lingerie set & 0.8122 & 0.0858 & 0.7264 & 59.68\% \\ \hline
        onlypuff Pocket Shirts for Women Casual & 1.0000 & 0.3250 & 0.6750 & 27.42\% \\ \hline
        Ekouaer Long Nightgown,Women's Loungewear Short Sleeve & 0.8415 & 0.1825 & 0.6590 & 33.87\% \\ \hline
        Lugz Women's Empire Hi Wvt Fashion Boot & 1.0000 & 0.3441 & 0.6559 & 24.19\% \\ \hline
        Dinosaur Pee Pee Teepee Wee & 1.0000 & 0.3447 & 0.6553 & 41.94\% \\ \hline
        10th birthday decorations for girl & 0.9558 & 0.3010 & 0.6548 & 19.35\% \\ \hline
        juDanzy kids knee high tube socks with grips & 1.0000 & 0.3539 & 0.6461 & 58.06\% \\ \hline
        Matching Delivery Robe and Swaddle Blanket & 1.0000 & 0.3571 & 0.6429 & 41.94\% \\ \hline
        UCGOU Bubble Mailers 7.25x12 Inch Teal 25 Pack & 0.9117 & 0.2840 & 0.6277 & 24.19\% \\ \hline
        Cicy Bell Women's Sunflower & 0.8891 & 0.2680 & 0.6211 & 62.90\% \\ \hline
        Canomo Lamp Light Kit Make a & 0.9157 & 0.2948 & 0.6209 & 19.35\% \\ \hline
        5L matte black rectangular trash can with soft close lid and anti-bag slip liner for bathroom or kitchen & 0.8923 & 0.2717 & 0.6206 & 50.00\% \\ \hline
        Women's UPF 50+ cotton linen bucket sun hat beige small & 1.0000 & 0.3837 & 0.6163 & 24.19\% \\ \hline
        Small breed wet dog food Hill's Science Diet Chicken \& Barley Recipe & 0.9904 & 0.3866 & 0.6038 & 40.32\% \\ \hline
        DKB Evian Jetted Whirlpool & 0.9083 & 0.3052 & 0.6031 & 27.42\% \\ \hline
        OtterBox Symmetry Disney Princess Mulan iPhone Xs iPhone X case & 0.9242 & 0.3240 & 0.6002 & 45.16\% \\ \hline
        Stars in the Desert book & 1.0000 & 0.4042 & 0.5958 & 58.06\% \\ \hline
        DC Collectibles Batman Arkham Origins & 0.9369 & 0.3439 & 0.5930 & 35.48\% \\ \hline
        fall sunflower pumpkin placemats set of 6 cotton linen washable table mats & 0.9813 & 0.3960 & 0.5853 & 27.42\% \\ \hline
        300 piece jigsaw puzzle Kitchen Memories by Steve Crisp & 1.0000 & 0.4179 & 0.5821 & 30.65\% \\ \hline
        Marvel Avengers Endgame Gauntlet T-Shirt & 0.9842 & 0.4026 & 0.5816 & 20.97\% \\ \hline
        Pahajim Women Fashion Purses Handbags Shoulder Tote Bags Top Handle Satchel & 1.0000 & 0.4189 & 0.5811 & 46.77\% \\ \hline
        Mai Puru Endo Mai's First Photo Collection & 0.6353 & 0.0547 & 0.5806 & 66.13\% \\ \hline
        girls princess dress up costume headband accessories & 0.8645 & 0.2881 & 0.5764 & 19.35\% \\ \hline
        Nanatang Badflower Logo Men's Long Sleeve Sweatshirt's & 0.9375 & 0.3619 & 0.5756 & 14.52\% \\ \hline
        HAPY SHOP 80 Pcs Silver Alligator Hair & 0.8525 & 0.2820 & 0.5705 & 45.16\% \\ \hline
        VERYKE L-Shaped sectional sofa chenille fabric golden legs living room & 0.9095 & 0.3406 & 0.5689 & 27.42\% \\ \hline
        ZINUS Owen Wood Platform & 1.0000 & 0.4333 & 0.5667 & 29.03\% \\ \hline
        Xperia 10 II Blue Light Cut Glass Film Asahi Japanese Ultra Thin Anti-Bubble Anti-Fingerprint & 0.6367 & 0.0737 & 0.5630 & 67.74\% \\ \hline
        Acer V6 V196LB 19" HD & 0.6797 & 0.1178 & 0.5619 & 50.00\% \\ \hline
        eyebrow tattoo stickers natural black dark brown & 0.5980 & 0.0393 & 0.5587 & 80.65\% \\ \hline
        buffalo plaid flannel pajama pants for women with pockets & 0.7937 & 0.2350 & 0.5587 & 35.48\% \\ \hline
        Women's Hotouch fringe vest faux suede 70s hippie sleeveless cardigan & 1.0000 & 0.4460 & 0.5540 & 35.48\% \\ \hline
        Microsoft Surface Laptop 2 & 0.9698 & 0.4175 & 0.5523 & 25.81\% \\ \hline
        men's sauna suit diet pants sportswear fat burner fitness wear exercise running weight loss black & 0.6477 & 0.0972 & 0.5505 & 59.68\% \\ \hline
        Fabri.YWL 1955 Ford Thunderbird Vintage Look Reproduction & 0.9639 & 0.4182 & 0.5457 & 37.10\% \\ \hline
        Figma Avengers Age of Ultron Iron Man Mark 43 Exclusive & 0.8974 & 0.3579 & 0.5395 & 20.97\% \\ \hline
        Under Armour Men's Jungle Rat Tactical & 0.9614 & 0.4224 & 0.5390 & 24.19\% \\ \hline
        Black Cast Iron Norfolk Door Latch 8" Tall Thumb Lock with Mounting Hardware & 0.9056 & 0.3707 & 0.5349 & 35.48\% \\ \hline
        CRAFTSMAN CMXZVBE38759 2-1/2 in. x 20 ft. POS-I-LOCK & 0.7560 & 0.2226 & 0.5334 & 58.06\% \\ \hline
        Columbia Women's Newton Ridge Waterproof Hiking Boot lightweight comfortable & 0.9567 & 0.4238 & 0.5329 & 17.74\% \\ \hline
        KATUMO Baby Crib Mobile Woodland Hot Air Balloons Birds Clouds Nursery Decoration & 1.0000 & 0.4707 & 0.5293 & 22.58\% \\ \hline
        Spider-Man Miles Morales Venom 3 Pack T-Shirts for kids & 0.7881 & 0.2589 & 0.5292 & 53.23\% \\ \hline
        double zipper smart key case genuine leather & 0.8514 & 0.3239 & 0.5275 & 14.52\% \\ \hline
        Alex Evenings womens Plus Size Midi Scoop Neck Shift & 0.8538 & 0.3269 & 0.5269 & 33.87\% \\ \hline
        cotton candy party kit 100 cones five flavors & 0.8898 & 0.3647 & 0.5251 & 25.81\% \\ \hline
        black wood electric guitar ornament & 1.0000 & 0.4852 & 0.5148 & 24.19\% \\ \hline
        Motorola Moto E6 Play 5.5" pantalla, 2GB RAM, 32GB almacenamiento, Android 9.0, Dual SIM, Gris & 0.9307 & 0.4170 & 0.5137 & 20.97\% \\ \hline
        Harley-Davidson Enjoy Ride Oval Embossed Tin Sign, 18 & 1.0000 & 0.4875 & 0.5125 & 24.19\% \\ \hline
        Quiksilver Men's Long Sleeve Rashguard UPF 50 Sun Protection Surf Shirt & 0.8611 & 0.3492 & 0.5119 & 16.13\% \\ \hline
        Eternal Nail Polish Set (BEACH WALK) & 0.9788 & 0.4681 & 0.5107 & 20.97\% \\ \hline
        CINRA 20PCS Disposable Tattoo Tubes Grips Mixed Sizes & 0.8786 & 0.3683 & 0.5103 & 27.42\% \\ \hline
        Dotty the Dalmatian plush mascot costume men's & 1.0000 & 0.4950 & 0.5050 & 29.03\% \\ \hline
        Levi's Women's High Waisted Mom Shorts & 1.0000 & 0.4954 & 0.5046 & 14.52\% \\ \hline
        HDMI to VGA DVI HDMI adapter 4 in 1 video converter for laptop monitor projector & 0.7962 & 0.2922 & 0.5040 & 32.26\% \\ \hline
        Nike Dri-Fit Men's Half Zip Golf Top & 0.8553 & 0.3514 & 0.5039 & 17.74\% \\ \hline
        MixMatchy Women's Striped Print Ribbed Knit Crop & 1.0000 & 0.4976 & 0.5024 & 12.90\% \\ \hline
        Dr. Martens Women's Shriver Hi Fashion Boot chunky sole & 0.8888 & 0.3906 & 0.4982 & 25.81\% \\ \hline
        Thankful Deluxe Fall Party Dinnerware Bundle with Dinner Plates, Dessert Plates, and Large Napkins & 0.9658 & 0.4710 & 0.4948 & 19.35\% \\ \hline
        Summer Clean Rinse Baby Bather (Gray) & 0.9713 & 0.4771 & 0.4942 & 25.81\% \\ \hline
        large black hair claw clips for thick hair women girls & 0.7120 & 0.2186 & 0.4934 & 27.42\% \\ \hline
        green camouflage RFID trifold canvas wallet for men with mini coin purse and front pocket for kids & 0.8033 & 0.3148 & 0.4885 & 32.26\% \\ \hline
        Find RAINBEAN foldable laptop table for bed with storage space and bamboo wood grain & 0.9647 & 0.4786 & 0.4861 & 19.35\% \\ \hline
        short black cosplay wig for men & 0.8409 & 0.3579 & 0.4830 & 20.97\% \\ \hline
        Detroit Axle - 4WD 8-Lug Front Wheel & 0.8406 & 0.3578 & 0.4828 & 24.19\% \\ \hline
        Apple Lime Green Deco Mesh Ribbon 10 inch for Christmas Wreath and Spring Decorations & 0.9260 & 0.4444 & 0.4816 & 24.19\% \\ \hline
        RJ-Sport No-Tie, Elastic Shoelaces, & 1.0000 & 0.5212 & 0.4788 & 22.58\% \\ \hline
        hypoallergenic baby bottle dishwashing liquid fragrance free & 0.9267 & 0.4495 & 0.4772 & 24.19\% \\ \hline
        AUGYMER stainless steel serrated bread knife 7.9 inches & 1.0000 & 0.5245 & 0.4755 & 20.97\% \\ \hline
        Kaenon Men's Polarized Sunglasses Burnet Full Coverage Matte Tortoise & 1.0000 & 0.5278 & 0.4722 & 20.97\% \\ \hline
        2-pack aluminum no firearms guns or weapons allowed sign 10x7 waterproof & 0.9922 & 0.5203 & 0.4719 & 25.81\% \\ \hline
        LED Candles, Ymenow Warm & 0.6399 & 0.1688 & 0.4711 & 37.10\% \\ \hline
        Ralph Lauren Meadow Lane Kaley King Comforter Blue Multi & 0.9902 & 0.5192 & 0.4710 & 19.35\% \\ \hline
        Portable RGB Gaming Mechanical Keyboard USB Type & 0.8618 & 0.3913 & 0.4705 & 17.74\% \\ \hline
        Coppertone Defend \& Care Sensitive Skin Sunscreen Lotion Broad & 0.8670 & 0.4012 & 0.4658 & 24.19\% \\ \hline
        Nike Kids' Grade School Zoom Air Pegasus & 0.9933 & 0.5282 & 0.4651 & 27.42\% \\ \hline
        Blue rectangular plastic tablecloth 54x108 inch for parties and outdoor & 0.9292 & 0.4659 & 0.4633 & 14.52\% \\ \hline
        Talking Products, Talking Tile & 0.6971 & 0.2342 & 0.4629 & 51.61\% \\ \hline
        Double sided rug tape for laminate flooring 2 inch 30 yards & 0.9273 & 0.4690 & 0.4583 & 19.35\% \\ \hline
        Nordic Ware Platinum Collection Anniversary Bundtlette & 0.8968 & 0.4404 & 0.4564 & 30.65\% \\ \hline
        glass essential oil diffuser ultrasonic aromatherapy humidifier wood natural & 0.9373 & 0.4842 & 0.4531 & 20.97\% \\ \hline
        purple ergonomic document copy holder with side arm & 0.8840 & 0.4311 & 0.4529 & 25.81\% \\ \hline
        iPhone 6 Plus replacement touch screen LCD panel with repair tools & 0.7864 & 0.3340 & 0.4524 & 17.74\% \\ \hline
        TRUE LINE Automotive Universal 7 Inch & 0.4998 & 0.0486 & 0.4512 & 77.42\% \\ \hline
        Kiss Tweetheart False Nails with accents and super hold adhesive & 1.0000 & 0.5496 & 0.4504 & 19.35\% \\ \hline
        Mayton wooden bunkie board slats Twin XL Beige & 1.0000 & 0.5506 & 0.4494 & 29.03\% \\ \hline
        16 oz clear plastic cups with flat lids set of 50 & 0.9556 & 0.5091 & 0.4465 & 11.29\% \\ \hline
        Benefit Boi ing Hydrating Concealer Light Medium & 1.0000 & 0.5560 & 0.4440 & 27.42\% \\ \hline
        Google Wi-Fi System Mesh & 1.0000 & 0.5565 & 0.4435 & 16.13\% \\ \hline
        YONEX AC1025P Tennis Badminton Grip & 0.4920 & 0.0488 & 0.4432 & 70.97\% \\ \hline
        ASICS Women's Gel-Venture 7 trail running shoes & 1.0000 & 0.5592 & 0.4408 & 32.26\% \\ \hline
        Blushing Cherry Blossom Fragrance Oil (60ml) For & 0.8977 & 0.4585 & 0.4392 & 19.35\% \\ \hline
        Ateco Pastry Leaf Tube Set White & 1.0000 & 0.5610 & 0.4390 & 24.19\% \\ \hline
        Garmin ECHOMAP Plus 73cv Ice & 0.9540 & 0.5151 & 0.4390 & 24.19\% \\ \hline
        7x5ft Black and Gold Balloons Backdrop for Birthday Party Photography & 0.7982 & 0.3599 & 0.4383 & 22.58\% \\ \hline
        5 Pcs Colorful Mini Silicone Whisks Stainless Steel Non Stick for Cooking Baking & 0.9337 & 0.4956 & 0.4381 & 17.74\% \\ \hline
        Custom Self Inking Rubber Stamp 4 Lines Extra Ink Pad A1848 & 0.8462 & 0.4123 & 0.4339 & 20.97\% \\ \hline
        Organic baby toddler short sleeve tight fit pajamas Lamaze & 0.7029 & 0.2693 & 0.4336 & 29.03\% \\ \hline
    \end{tabular}}
    \caption{Per Query Mean NDCG Scores and Variability on first 100 queries. For each query, we measure using NDCG@10 the min, mean, gap between mean and max, and the percentage of systems that have an NDCG@10 of 0, denoting a completely irrelevant retrieval set}
    \label{tab:per-query-pt1}
\end{table}

\begin{table}[]
    \centering
    \tiny
    \scalebox{0.8}{
    \begin{tabular}{|l|l|l|l|l|}
    \hline
        Query & Max & Mean & Gap Max vs Mean & Systems with NDCG@10 = 0 \\ \hline
        Microsoft Surface Pro Signature Type Cover Platinum FFP-00141 for Surface Pro 7 & 0.9721 & 0.5385 & 0.4336 & 17.74\% \\ \hline
        Kids sandwich cutter and sealer set with star, heart, and circle shapes for lunchbox and bento box & 0.9181 & 0.4863 & 0.4318 & 22.58\% \\ \hline
        ergonomic stool adjustable height footrest ring parquet wheels & 0.6176 & 0.1858 & 0.4318 & 24.19\% \\ \hline
        Onvian Bike Alarm with Remote, Wireless & 1.0000 & 0.5709 & 0.4291 & 22.58\% \\ \hline
        acrylic lollipop holder cake pop stand with sticks, bags, and twist ties & 0.8547 & 0.4272 & 0.4275 & 16.13\% \\ \hline
        90 degree right angled Micro USB 2.0 Male to Female extension cable 50cm for tablet phone & 0.9359 & 0.5108 & 0.4251 & 14.52\% \\ \hline
        Goodman 4 Ton 14 & 0.7710 & 0.3463 & 0.4247 & 24.19\% \\ \hline
        OTTERBOX COMMUTER SERIES iPhone 11 Pro case with PopSockets PopGrip & 0.8303 & 0.4067 & 0.4236 & 29.03\% \\ \hline
        Probrico flat black cabinet pulls 3 inch hole centers T bar handle kitchen dresser 10 pack & 0.8018 & 0.3829 & 0.4190 & 22.58\% \\ \hline
        Madagascar Bourbon Planifolia Grade A Vanilla Beans 5-6 inches & 0.8182 & 0.4013 & 0.4169 & 27.42\% \\ \hline
        Nerf Mega Accustrike Dart Refill Combat Blaster & 0.8542 & 0.4395 & 0.4147 & 17.74\% \\ \hline
        100\% Cotton Throw Blanket for Couch Sofa Bed Outdoors Hypoallergenic 83"x70" Brown & 0.8888 & 0.4753 & 0.4135 & 8.06\% \\ \hline
        Zoostliss Crimping Tool for Coaxial Cable RG6 RG59 with Blue F Connector & 0.9872 & 0.5770 & 0.4102 & 17.74\% \\ \hline
        adidas ladies climalite short-sleeve polo & 0.8080 & 0.3980 & 0.4100 & 16.13\% \\ \hline
        Hofdeco nautical indoor outdoor pillow cover navy blue compass anchor 12x20 set of 2 & 0.8266 & 0.4170 & 0.4096 & 20.97\% \\ \hline
        Hurley Men's Printed Backpack Light Carbon & 0.9072 & 0.4995 & 0.4077 & 27.42\% \\ \hline
        Giorgio Armani Code Colonia men's Eau de Toilette Spray 2.5 oz & 0.9571 & 0.5504 & 0.4067 & 20.97\% \\ \hline
        Extra Large Moist Heating Pad & 1.0000 & 0.5952 & 0.4048 & 20.97\% \\ \hline
        HOBO Vintage Euro Slide Credit Card Holder Wallet & 0.7783 & 0.3753 & 0.4030 & 20.97\% \\ \hline
        Gueray Portable CD Player, & 0.9758 & 0.5729 & 0.4029 & 16.13\% \\ \hline
        HDMI to RCA converter for TV, Roku, Fire Stick, DVD, Blu-ray player & 1.0000 & 0.5973 & 0.4027 & 12.90\% \\ \hline
        Ombre highlight short bob human hair wig brown to blonde Brazilian straight remy for black women & 0.9140 & 0.5142 & 0.3998 & 12.90\% \\ \hline
        Muga Black Felt Letter Board 12x12 inch with 485 Precut Letters and Stand & 0.7829 & 0.3846 & 0.3983 & 19.35\% \\ \hline
        Deflecto 5-bin horizontal tilt bin storage system black & 0.9789 & 0.5811 & 0.3978 & 17.74\% \\ \hline
        Iridescent Metal Lighter Case for BIC Lighters, Lighter & 0.9098 & 0.5131 & 0.3967 & 20.97\% \\ \hline
        Jill \& Joey Maternity Belt Belly Band Medium Beige Pregnancy Support & 0.8930 & 0.4980 & 0.3950 & 20.97\% \\ \hline
        Calvin Klein Women's Invisibles Hipster Panty & 1.0000 & 0.6086 & 0.3914 & 19.35\% \\ \hline
        KOKUYO Campus Notebook B5 Dotted A-Ruled 5-Colors 6 Pack & 0.9047 & 0.5154 & 0.3893 & 24.19\% \\ \hline
        magnetic dart board for kids boys girls gifts & 1.0000 & 0.6108 & 0.3892 & 22.58\% \\ \hline
        Rokinon Cine DS Lens Kit Micro Four Thirds 16mm 35mm 50mm 85mm & 0.9597 & 0.5726 & 0.3871 & 22.58\% \\ \hline
        Zealer 1800pcs Crystals AB Nail Art Rhinestones Flatback Glass Charms for Nails Decoration Eye Makeup Clothes Shoes Mix Sizes & 0.7793 & 0.3936 & 0.3857 & 24.19\% \\ \hline
        Perricone MD Neuropeptide Facial Conformer & 0.8735 & 0.4879 & 0.3856 & 22.58\% \\ \hline
        Soda Pop Can Covers Made in USA BPA-Free Retains Fizz & 0.9023 & 0.5233 & 0.3790 & 17.74\% \\ \hline
        Baby Toddler Girls Long Sleeve & 1.0000 & 0.6216 & 0.3784 & 9.68\% \\ \hline
        Blue pet grooming gloves for hair removal and massage & 0.9742 & 0.5976 & 0.3766 & 19.35\% \\ \hline
        AnnTec LED Candle Light, LED Glass, & 0.4367 & 0.0602 & 0.3765 & 67.74\% \\ \hline
        Osprey Daylite Shoulder Sling for daily essentials and quick hikes & 0.8261 & 0.4508 & 0.3753 & 12.90\% \\ \hline
        Tminnov Baby Diaper Caddy & 0.8567 & 0.4832 & 0.3735 & 12.90\% \\ \hline
        NFL Women's OTS Fleece Hoodie & 0.9003 & 0.5308 & 0.3695 & 17.74\% \\ \hline
        39 watt halogen PAR20 clear medium base bulb & 0.7402 & 0.3713 & 0.3689 & 17.74\% \\ \hline
        Daiwa Liberty Club Short Swing & 0.3904 & 0.0224 & 0.3680 & 91.94\% \\ \hline
        4 ton hydraulic low profile floor jack with dual piston quick lift pump & 0.9062 & 0.5384 & 0.3678 & 20.97\% \\ \hline
        universal knife and tool sharpener with adjustable angle guides & 0.6768 & 0.3095 & 0.3673 & 22.58\% \\ \hline
        OTC 4842 Heavy Duty Valve Spring Compressor & 1.0000 & 0.6335 & 0.3665 & 19.35\% \\ \hline
        Spencer 50'/15M Logger Tape Refill & 0.9121 & 0.5463 & 0.3658 & 19.35\% \\ \hline
        Toozey Dog Pooper Scooper, Upgraded Adjustable Long & 1.0000 & 0.6383 & 0.3617 & 16.13\% \\ \hline
        Lovelyshop Blue Gems Rhinestone & 0.4991 & 0.1385 & 0.3606 & 22.58\% \\ \hline
        sparkly rhinestone mesh face mask for women masquerade party glitter bling Christmas & 0.9120 & 0.5517 & 0.3603 & 12.90\% \\ \hline
        Enameled Cast Iron Dutch Oven 6.5 quart Olive Green & 0.9013 & 0.5429 & 0.3584 & 16.13\% \\ \hline
        Huffy 12V Battery Powered Ride On ATV for kids aged 2-5 & 0.7172 & 0.3600 & 0.3572 & 24.19\% \\ \hline
        artificial eucalyptus plant for wedding jungle theme party home decor & 1.0000 & 0.6433 & 0.3567 & 19.35\% \\ \hline
        Klein Tools Mini Tube Cutter for copper and aluminum tubing & 0.7522 & 0.3980 & 0.3542 & 25.81\% \\ \hline
        An Untamed Heart Red River of the North book & 1.0000 & 0.6463 & 0.3537 & 24.19\% \\ \hline
        New American citizen mug US flag coffee cup gift & 0.9053 & 0.5531 & 0.3522 & 12.90\% \\ \hline
        Ernie Ball Custom Gauge 11 Nickel Guitar String 6 Pack & 0.8578 & 0.5087 & 0.3491 & 24.19\% \\ \hline
        Bobbi Brown Lip Color Rum Raisin 3.4g & 1.0000 & 0.6513 & 0.3487 & 19.35\% \\ \hline
        21x21 beige sun shade sail patio UV protection outdoor backyard & 0.7790 & 0.4350 & 0.3440 & 12.90\% \\ \hline
        Carhartt Men's Heavyweight Short-Sleeve Pocket T-Shirt & 1.0000 & 0.6572 & 0.3428 & 16.13\% \\ \hline
        Hokkaido Deer Lotion CICA Ceramide Vitamin C 5.1 fl oz & 0.4177 & 0.0752 & 0.3425 & 56.45\% \\ \hline
        Rise Pea Protein Bar, & 0.9030 & 0.5638 & 0.3393 & 19.35\% \\ \hline
        Hunter Watson Indoor ceiling fan with LED light and pull chain control, New Bronze finish & 0.9504 & 0.6115 & 0.3389 & 16.13\% \\ \hline
        Belkin QODE Ultimate Pro Keyboard Case for iPad Air 2 White & 0.9291 & 0.5909 & 0.3382 & 17.74\% \\ \hline
        The Fifth Agreement book self-mastery Toltec Wisdom & 0.9427 & 0.6060 & 0.3367 & 14.52\% \\ \hline
        Echo Dot Wall Mount Holder Black AhaStyle & 0.9878 & 0.6540 & 0.3338 & 4.84\% \\ \hline
        Soft Spiked Light Up Bracelets with Flashing Blinking LED Lights & 0.7561 & 0.4228 & 0.3333 & 14.52\% \\ \hline
        migraine relief hat hot cold therapy reusable MarkGifts & 0.9336 & 0.6005 & 0.3331 & 20.97\% \\ \hline
        cordless portable blender rechargeable battery smoothie mixer travel & 1.0000 & 0.6693 & 0.3307 & 19.35\% \\ \hline
        LYXOTO Baby Hair Clips set of 10 bow knot cute stylish birthday gift & 0.7265 & 0.3989 & 0.3276 & 11.29\% \\ \hline
        Plaskidy plastic forks for kids set of 16 BPA free dishwasher safe & 0.7287 & 0.4106 & 0.3181 & 12.90\% \\ \hline
        Nantucket Neighbors (Nantucket Beach Plum Cove Book 2) & 1.0000 & 0.6838 & 0.3162 & 16.13\% \\ \hline
        Physicians Formula Highlighter Makeup Powder Mineral Glow Pearls, Light Bronze & 0.9623 & 0.6466 & 0.3157 & 20.97\% \\ \hline
        Organic Kabuli Chana 4 lb White & 0.7598 & 0.4503 & 0.3095 & 20.97\% \\ \hline
        PEARL IZUMI Elite Thermal Arm Warmer for cool weather cycling with water-shedding fleece and unisex sizing & 0.9928 & 0.6874 & 0.3054 & 19.35\% \\ \hline
        GE 6-Device Backlit Universal Remote Control for Samsung, Vizio, LG, Sony, Sharp, Roku, Apple TV, Blu-Ray, DVD, Master Volume Control, Rose Gold & 1.0000 & 0.6963 & 0.3037 & 19.35\% \\ \hline
        SPLOTY Tire Inflator Air Compressor Portable 12V & 1.0000 & 0.6979 & 0.3021 & 19.35\% \\ \hline
        VAKA Luminous Light Up Quad Roller Skate & 0.6643 & 0.3624 & 0.3019 & 19.35\% \\ \hline
        Black+Decker 5-Cup Coffeemaker Dust Proof Cover & 0.6907 & 0.3924 & 0.2983 & 20.97\% \\ \hline
        Celestial Seasonings Honey Vanilla Chamomile Herbal Tea 20 Count & 0.9059 & 0.6086 & 0.2973 & 17.74\% \\ \hline
        large washable donut dog bed with calming cuddler and head support & 0.8315 & 0.5356 & 0.2959 & 12.90\% \\ \hline
        professional hair cutting scissors extremely sharp blades 6 inch barber scissors set for men and women Fagaci & 0.7568 & 0.4615 & 0.2953 & 17.74\% \\ \hline
        Stoplight Sleep Enhancing Alarm Clock for Kids Train Car Clock & 0.7001 & 0.4072 & 0.2929 & 17.74\% \\ \hline
        McCormick Coconut Extract 2 fl oz gluten-free non-GMO & 0.6143 & 0.3433 & 0.2710 & 19.35\% \\ \hline
        XR Extinction Rebellion Rebel For Life T-Shirt & 0.9449 & 0.6863 & 0.2586 & 11.29\% \\ \hline
        Dogfish 500GB Msata Internal SSD & 0.2658 & 0.0126 & 0.2532 & 91.94\% \\ \hline
        Car glass coating agent with shampoo conditioner and water repellent for glossy finish & 0.2038 & 0.0398 & 0.1640 & 53.23\% \\ \hline
    \end{tabular}}
    \caption{Per Query Mean NDCG Scores and Variability on remaining 82 queries. For each query, we measure using NDCG@10 the min, mean, gap between mean and max, and the percentage of systems that have an NDCG@10 of 0, denoting a completely irrelevant retrieval set}
    \label{tab:per-query-pt2}
\end{table}
\section{Impact of Finetuning on single stage Language Model Based Retrieval}
\begin{table}[!ht]
    \centering
    \begin{tabular}{|l|l|l|l|l|}
    \hline
        Run & Collection Type & Zero Shot & Fine-tune & Impact \\ \hline
        all-MiniLM-L12-v2 & Simple & 0.4952 & 0.4763 & -0.0189 \\ \hline
        all-MiniLM-L12-v2 & Metadata & 0.5328 & 0.4681 & -0.0647 \\ \hline
        all-miniLM-L6-v2 & Simple & 0.5895 & 0.4591 & -0.1303 \\ \hline
        all-miniLM-L6-v2 & Metadata & 0.5937 & 0.4673 & -0.1264 \\ \hline
        all-mpnet-base-v2 & Simple & 0.5895 & 0.4090 & -0.1804 \\ \hline
        all-mpnet-base-v2 & Metadata & 0.5937 & 0.4089 & -0.1848 \\ \hline
        bge-base-en & Simple & 0.2948 & 0.4064 & 0.1116 \\ \hline
        bge-base-en & Metadata & 0.3396 & 0.4237 & 0.0841 \\ \hline
        bge-large-en & Metadata & 0.0323 & 0.0021 & -0.0302 \\ \hline
        bge-small-en & Simple & 0.3680 & 0.4379 & 0.0699 \\ \hline
        e5-large-v2 & Simple & 0.4998 & 0.0015 & -0.4983 \\ \hline
        e5-base-v2 & Metadata & 0.0936 & 0.4242 & 0.3306 \\ \hline
        e5-small-v2 & Metadata & 0.0142 & 0.5119 & 0.4977 \\ \hline
        gte-base & Simple & 0.1493 & 0.4777 & 0.3284 \\ \hline
        gte-base & Metadata & 0.0604 & 0.5009 & 0.4405 \\ \hline
        gte-large & Simple & 0.3930 & 0.0015 & -0.3915 \\ \hline
        gte-large & Metadata & 0.4501 & 0.0011 & -0.4490 \\ \hline
        gte-small & Simple & 0.6612 & 0.4901 & -0.1711 \\ \hline
        gte-small & Metadata & 0.6647 & 0.5168 & -0.1479 \\ \hline
        bert-base & Metadata & 0.4165 & 0.0352 & -0.3813 \\ \hline
    \end{tabular}
    \caption{Impact on NDCG@10 of fine-tuning across baseline runs.}
    \label{tab:finetune-zero-n}
\end{table}
\begin{table}[!ht]
    \centering
    \begin{tabular}{|l|l|l|l|l|}
    \hline
        Run & Collection Type & Zero Shot & Fine-tune & Impact \\ \hline
        all-MiniLM-L12-v2 & Simple & 0.6099 & 0.5351 & -0.0747 \\ \hline
        all-MiniLM-L12-v2 & Metadata & 0.6415 & 0.5581 & -0.0833 \\ \hline
        all-miniLM-L6-v2 & Simple & 0.6348 & 0.5429 & -0.0920 \\ \hline
        all-miniLM-L6-v2 & Metadata & 0.6512 & 0.5528 & -0.0984 \\ \hline
        all-mpnet-base-v2 & Simple & 0.6348 & 0.4747 & -0.1602 \\ \hline
        all-mpnet-base-v2 & Metadata & 0.6512 & 0.5006 & -0.1506 \\ \hline
        bge-base-en & Simple & 0.2664 & 0.4728 & 0.2064 \\ \hline
        bge-base-en & Metadata & 0.4301 & 0.4955 & 0.0654 \\ \hline
        bge-large-en & Metadata & 0.0314 & 0.0000 & -0.0314 \\ \hline
        bge-small-en & Simple & 0.4188 & 0.5080 & 0.0893 \\ \hline
        e5-large-v2 & Simple & 0.5537 & 0.0000 & -0.5537 \\ \hline
        e5-base-v2 & Metadata & 0.1021 & 0.5148 & 0.4127 \\ \hline
        e5-small-v2 & Metadata & 0.0126 & 0.6096 & 0.5970 \\ \hline
        gte-base & Simple & 0.0965 & 0.5676 & 0.4711 \\ \hline
        gte-base & Metadata & 0.0940 & 0.5895 & 0.4956 \\ \hline
        gte-large & Simple & 0.4294 & 0.0001 & -0.4293 \\ \hline
        gte-large & Metadata & 0.4978 & 0.0000 & -0.4978 \\ \hline
        gte-small & Simple & 0.7372 & 0.5692 & -0.1679 \\ \hline
        gte-small & Metadata & 0.7416 & 0.5859 & -0.1557 \\ \hline
        bert-base & Metadata & 0.5208 & 0.0374 & -0.4834 \\ \hline
    \end{tabular}
    \caption{Impact on Recall@100 of fine-tuning across baseline runs.}
    \label{tab:finetune-zero-r}
\end{table}
\section{Stratifying based on query length}
\begin{table}[!htb]
    \centering
    \small
    \begin{tabular}{|l|l|l|l|}
    \hline
        Run & NDCG @10 (All) & NDCG@10 (Short) & NDCG@10 (Long) \\ \hline
        f\_splade\_bm25 & 0.7505 & 0.7407 & 0.7578 \\ \hline
        f\_splade\_clip\_bm25 & 0.7327 & 0.7078 & 0.7511 \\ \hline
        cfdaclip\_ER\_B & 0.7256 & 0.7119 & 0.7358 \\ \hline
        cfdaclip\_MR\_B & 0.7251 & 0.7108 & 0.7357 \\ \hline
        cfdaclip\_MR\_A & 0.7257 & 0.7122 & 0.7357 \\ \hline
        cfdaclip\_ER\_A & 0.7252 & 0.7111 & 0.7356 \\ \hline
        JBNU-C & 0.7251 & 0.7130 & 0.7341 \\ \hline
        JBNU-1 & 0.6531 & 0.6060 & 0.6878 \\ \hline
        simple-gte-small-zero-shot & 0.6612 & 0.6253 & 0.6877 \\ \hline
        metadata-enhanced-gte-small-zero-shot & 0.6647 & 0.6343 & 0.6872 \\ \hline
        JBNU-2 & 0.6583 & 0.6238 & 0.6838 \\ \hline
        BM25-pyserini-simple-collection & 0.6540 & 0.6184 & 0.6803 \\ \hline
        BM25-pyserini-metadata-collection & 0.6408 & 0.6043 & 0.6643 \\ \hline
        f\_gpt\_rerank & 0.6225 & 0.6162 & 0.6273 \\ \hline
        JBNU-A & 0.5989 & 0.5616 & 0.6265 \\ \hline
        metadata-enhanced-all-mpnet-base-v2-zero-shot & 0.5937 & 0.5586 & 0.6197 \\ \hline
        JBNU-B & 0.5763 & 0.5294 & 0.6110 \\ \hline
        metadata-enhanced-all-MiniLM-L12-v2-zero-shot & 0.5660 & 0.5072 & 0.6093 \\ \hline
        r\_gpt3d5\_turbo & 0.5950 & 0.5766 & 0.6087 \\ \hline
        simple-all-mpnet-base-v2-zero-shot & 0.5895 & 0.5687 & 0.6048 \\ \hline
        simple-all-MiniLM-L12-v2-zero-shot & 0.5288 & 0.4721 & 0.5707 \\ \hline
        metadata-enhanced-trec-product-search-gte-small & 0.5168 & 0.4457 & 0.5693 \\ \hline
        metadata-enhanced-all-MiniLM-L6-v2-zero-shot & 0.5328 & 0.4938 & 0.5616 \\ \hline
        simple-e5-large-zero-shot & 0.4998 & 0.4230 & 0.5565 \\ \hline
        metadata-enhanced-trec-product-search-e5-small-v2 & 0.5119 & 0.4583 & 0.5514 \\ \hline
        metadata-enhanced-trec-product-search-gte-base & 0.5009 & 0.4471 & 0.5407 \\ \hline
        simple-trec-product-search-gte-small & 0.4901 & 0.4535 & 0.5171 \\ \hline
        metadata-enhanced-trec-product-search-bge-small-en & 0.4721 & 0.4192 & 0.5112 \\ \hline
        simple-all-MiniLM-L6-v2-zero-shot & 0.4952 & 0.4746 & 0.5104 \\ \hline
        metadata-trec-product-search-all-miniLM-L12-v2 & 0.4681 & 0.4193 & 0.5042 \\ \hline
        simple-trec-product-search-gte-base & 0.4777 & 0.4430 & 0.5032 \\ \hline
        metadata-trec-product-search-all-miniLM-L6-v2 & 0.4673 & 0.4427 & 0.4856 \\ \hline
        simple-trec-product-search-all-miniLM-L12-v2 & 0.4763 & 0.4650 & 0.4847 \\ \hline
        simple-trec-product-search-all-miniLM-L6-v2 & 0.4591 & 0.4444 & 0.4700 \\ \hline
        metadata-enhanced-trec-product-search-e5-base-v2 & 0.4242 & 0.3698 & 0.4644 \\ \hline
        metadata-enhanced-gte-large-zero-shot & 0.4501 & 0.4361 & 0.4604 \\ \hline
        metadata-enhanced-trec-product-search-bge-base-en & 0.4237 & 0.3847 & 0.4525 \\ \hline
        simple-trec-product-search-bge-small-en & 0.4379 & 0.4267 & 0.4462 \\ \hline
        search-dpr-bert-base & 0.4272 & 0.4029 & 0.4452 \\ \hline
        metadata-enhanced-trec-product-search-dpr-bert & 0.4165 & 0.3805 & 0.4430 \\ \hline
        metadata-trec-product-search-all-mpnet-base-v2 & 0.4089 & 0.3784 & 0.4315 \\ \hline
        simple-trec-product-search-all-mpnet-base-v2 & 0.4090 & 0.3851 & 0.4267 \\ \hline
        simple-trec-product-search-bge-base-en & 0.4064 & 0.3932 & 0.4161 \\ \hline
        simple-gte-large-zero-shot & 0.3930 & 0.3647 & 0.4140 \\ \hline
        simple-bge-small-zero-shot & 0.3680 & 0.3378 & 0.3903 \\ \hline
        metadata-enhanced-bge-base-en-zero-shot & 0.3396 & 0.2902 & 0.3761 \\ \hline
        simple-bge-base-zero-shot & 0.2948 & 0.2923 & 0.2967 \\ \hline
        simple-gte-base-zero-shot & 0.1493 & 0.1696 & 0.1342 \\ \hline
        simple-bge-large-zero-shot & 0.1486 & 0.1805 & 0.1250 \\ \hline
        simple-e5-base-zero-shot & 0.1168 & 0.1350 & 0.1033 \\ \hline
        metadata-enhanced-e5-base-v2-zero-shot & 0.0936 & 0.1127 & 0.0795 \\ \hline
        metadata-enhanced-gte-base-zero-shot & 0.0604 & 0.0580 & 0.0621 \\ \hline
        simple-bert-base-uncased-zero-shot & 0.0352 & 0.0417 & 0.0305 \\ \hline
        simple-contriever-base-zero-shot & 0.0159 & 0.0044 & 0.0244 \\ \hline
        metadata-enhanced-e5-small-v2-zero-shot & 0.0142 & 0.0046 & 0.0212 \\ \hline
        metadata-enhanced-bge-large-en-zero-shot & 0.0323 & 0.0495 & 0.0197 \\ \hline
        simple-e5-small-zero-shot & 0.0113 & 0.0046 & 0.0163 \\ \hline
        metadata-enhanced-contriever-base-msmarco & 0.0081 & 0.0032 & 0.0116 \\ \hline
        metadata-enhanced-trec-product-search-bge-large-en & 0.0021 & 0.0024 & 0.0019 \\ \hline
        simple-trec-product-search-gte-large & 0.0015 & 0.0022 & 0.0010 \\ \hline
        metadata-enhanced-trec-product-search-gte-large & 0.0011 & 0.0011 & 0.0010 \\ \hline
        metadata-enhanced-trec-product-search-e5-large-v2 & 0.0015 & 0.0023 & 0.0009 \\ \hline
    \end{tabular}
    \caption{Mean NDCG@10 performance of systems with full sample of queries (182), short queries (81), and long queries (101) across various systems.}
    \label{tab:query-length}
\end{table}
\end{document}